\documentclass[10pt,journal,compsoc,twoside]{IEEEtran}

\usepackage[T1]{fontenc}
\usepackage{graphicx}
\usepackage{wrapfig}
\usepackage{float}
\usepackage{graphicx,fancyvrb}
\usepackage{epstopdf}
\usepackage{amsmath,amssymb,amsfonts}
\usepackage{bm}
\usepackage[linesnumbered,boxed,ruled,commentsnumbered]{algorithm2e} 
\usepackage{textcomp}
\usepackage{array,booktabs,makecell}

\usepackage{subfigure}
\usepackage{enumerate}
\usepackage{enumitem}
\usepackage{float}
\usepackage[backref]{hyperref} 
\hypersetup{
	hidelinks
}
\DeclareMathAlphabet\mathbfcal{OMS}{cmsy}{b}{n}
\usepackage{hyperref}
\AppendGraphicsExtensions{.tif}

\SetKwInOut{Input}{\textbf{Input}}
\SetKwInOut{Output}{\textbf{Output}}
\SetKw{Continue}{continue}
\SetKw{Break}{break}

\usepackage{tcolorbox}
\usepackage{varwidth}
\usepackage{efbox}
\usepackage{adjustbox}
\usepackage{framed}
\usepackage{multirow}
\hypersetup{
	colorlinks=true,   
	linkcolor=red,  
	citecolor=black,   
	urlcolor=black,
}

\usepackage{multirow}
\usepackage{wrapfig}
\usepackage{threeparttable}

\usepackage{etoolbox}

\makeatletter

\long\def\@IEEEtitleabstractindextextbox#1{\parbox{0.922\textwidth}{#1}}

\pretocmd\@bibitem{\csname keycolor#1\endcsname}{}{\fail}
\newcommand\citecolor[3][1]{\@namedef{keycolor#3}{\hspace*{-\labelwidth}\hspace*{-\labelsep}{\color{#2}\rule[-0.3em]{\dimexpr\linewidth+\labelwidth+\labelsep\relax}{#1\baselineskip}}\vspace*{\itemsep}\vspace*{-#1\baselineskip}}}

\makeatother

\ifCLASSOPTIONcompsoc
  \usepackage[nocompress]{cite}
\else
  \usepackage{cite}
\fi

\usepackage[capitalise,noabbrev]{cleveref}
\usepackage{color}
\usepackage[normalem]{ulem}

\usepackage{acronym}
\acrodef{ML}{Machine Learning}
\acrodef{AI}{Artificial Intelligence}
\acrodef{RSS}{Received Signal Strength}
\acrodef{KPI}{Key Performance Indicator}
\acrodef{QoS}{Quality of Service}
\acrodef{SPM}{Standard Propagation Model}
\acrodef{SUI}{Stanford University Interim}
\acrodef{ANN}{Artificial Neural Network}
\acrodef{UE}{User Equipment}
\acrodef{BS}{Base Station}
\acrodef{DTM}{Digital Terrain Model}
\acrodef{DHM}{Digital Height Model}
\acrodef{DLU}{Digital Land Use Map}
\acrodef{LoS}{Line of Sight}
\acrodef{NLoS}{Non Line of Sight}
\acrodef{k-NN}{k-Nearest Neighbors}
\acrodef{DT}{Decision Tree}
\acrodef{RMSE}{Root Mean Square Error}
\acrodef{GBDT}{Gradient Boosting Decision Trees}
\acrodef{AdaBoost}{Adaptive Boosting}
\acrodef{XGBoost}{Extreme Gradient Boosting}
\acrodef{LightGBM}{Light Gradient Boosting Machine}
\acrodef{CatBoost}{Categorical Boosting}
\acrodef{DNN}{Deep Neural Network}
\acrodef{LIME}{Local Interpretable Model-Agnostic Explanations}
\acrodef{SHAP}{SHapley Additive exPlanations}
\acrodef{MDT}{Minimization of Drive Tests}
\acrodef{$R^2$}{coefficient of determination}
\acrodef{URLLC}{Ultra-Reliable Low-Latency Communication}
\acrodef{UAV}{Unmanned Aerial Vehicle}
\acrodef{RIS}{Reconfigurable Intelligent Surfaces}

\begin{document}

\title{Relative Attention-based One-Class Adversarial Autoencoder for Continuous Authentication of Smartphone Users}

%
%

\author{
	Mingming Hu,
	Kun Zhang,
	Ruibang You
	and Bibo Tu
	\IEEEcompsocitemizethanks{
		\IEEEcompsocthanksitem{
			This work was supported by the National Key Research and Development Program of China under grant No. 2023YFB3106303.
			
			\emph{(Corresponding author: Ding Wang and Bibo Tu).}}
		
		\IEEEcompsocthanksitem Mingming Hu is with Institute of Information Engineering, Chinese Academy of Sciences, and School of Cyber Security, University of Chinese Academy of Sciences, and also with the Yellow River Henan Bureau, e-mail: (14120390@bjtu.edu.cn).
		
		\IEEEcompsocthanksitem Kun Zhang, Ruibang You and Bibo Tu are with Institute of Information Engineering, Chinese Academy of Sciences, and School of Cyber Security, University of Chinese Academy of Sciences, e-mail: (\{majunchao, zhangkun, tubibo\}@iie.ac.cn).
	}
}


\IEEEtitleabstractindextext{

\begin{abstract}
Behavioral biometrics-based continuous authentication is a promising authentication scheme that uses behavioral biometrics recorded by built-in sensors to authenticate users throughout the session. However, current continuous authentication methods have some limitations: 1) behavioral biometrics from impostors are needed to train continuous authentication models. Since the distribution of negative samples from diverse attackers are unknown, it is difficult to solve in real-world scenarios; 2) most deep learning-based continuous authentication methods need to train two models to improve authentication performance. A deep learning model for deep feature extraction, and a machine learning-based classifier for classification; 3) weak capability of capturing users’ behavioral patterns leads to poor authentication performance. To solve these issues, we propose a relative attention-based one-class adversarial autoencoder for continuous authentication task. First, we propose a one-class adversarial autoencoder to learn latent representations of legitimate users' behavioral patterns, which is trained only with legitimate users' behavioral biometrics. Second, we present the relative attention layer to capture richer contextual semantic representation of users' behavioral patterns, which modifies the standard self-attention mechanism by using convolutional projection to perform the attention maps. Experimental results demonstrate that we achieve superior performance with equal error rate (EER) of 1.05\%, 1.09\%, and 1.08\% on three datasets.

\end{abstract}


}

\maketitle

\IEEEdisplaynontitleabstractindextext

\IEEEpeerreviewmaketitle

\acresetall

\section{Introduction}
\label{introduction}

With the improvement of smartphones' processing power and storage capacity, smartphones have become an indispensable tool in our daily lives. Users usually store a large amount of sensitive personal information and privacy information on their smartphones. Therefore, the leakage of personal privacy information on smartphones has aroused more and more people’s concerns. Once personal privacy information is leaked, it could have a major negative impact on individuals and the public. To prevent the personal privacy information on the smartphone from being accessed and obtained by unauthorized users, an authentication method that can effectively authenticate users who access the smartphone becomes more critical. Although the current authentication methods can perform identity authentication when the user accesses the smartphone, such as personal identification number (PIN), passwords, graphical passwords, fingerprint recognition, and face recognition, they only perform one-time authentication when smartphone users log in. Once an unauthorized user controls the smartphone after the smartphone is authenticated, privacy information is accessible and available until the smartphone logs out. Therefore, there is a great need for an authentication method that can continuously authenticate smartphone users throughout their session to enhance security.

Among various solutions, behavioral biometrics-based continuous authentication stands out as a promising solution. It utilizes sensory data from built-in smartphone sensors to authenticate smartphone users, measuring their behavioral patterns during interaction with the device. Compared with typical one-time authentication methods, continuous authentication based on behavioral biometrics has the following advantages: 1) no additional hardware support is required to obtain biometric data that can represent user's behavioral patterns; 2) the acquisition of sensor data does not require root access privilege; 3) smartphone user does not need to participate in the authentication process; 4) it can authenticate the user's identity throughout the session.

Although current methods based on traditional machine learning or deep learning have made exciting progress, they still suffer some limitations. First, sensory data from impostors (negative samples) is needed to train the continuous authentication model, whether for binary classification or multi-classification \cite{wang2021framework, abuhamad2020autosen, centeno2018mobile, li2020scanet, el2021implicit, zhu2020espialcog, giorgi2021using, hernandez2021smartcampp, li2023searchauth, li2024snnauth}, since the distribution of negative training data from diverse attackers are unknown, it is a difficult problem to solve in a real-world scenario. Besides, sharing of other smartphone users' biometric data (negative samples) may lead to biometric data leakage. Abuhamad et al. \cite{abuhamad2020autosen} proposed a long short-term memory (LSTM) architecture to capture smartphone users' behavioral patterns from sensory data. When training the model, in addition to selecting the sensory data from the legitimate user as positive samples, and they also use sensory data from other users as impostors' biometric data. Li et al. \cite{li2020scanet} proposed a continuous authentication system based on two-stream convolutional neural network. In the training phase, the two-stream convolutional neural network model is trained as a multi-classification task. Thus continuous authentication task should be considered as one-class learning or novelty detection task, as \cite{sitova2016hmog, shen2017performance, shen2022mmauth}. However, these continuous authentication methods cannot extract deep features from behavioral biometrics that can characterize smartphone user's behavioral patterns, resulting in unsatisfactory authentication performance. Second, the continuous authentication methods they proposed require training two models \cite{li2021deffusion, li2021cnn, centeno2018mobile, zhu2020hybrid, li2020scanet}, which require a significant amount of time and storage space consumption. Li et al. \cite{li2021cnn} employed a wasserstein generative adversarial network (CWGAN) to generate additional samples to augment the original training samples. In the training phase, they apply a convolutional neural network architecture to learning valuable deep features from sensor data, and the extracted deep features are fed to the one-class support vector machine (SVM) to classify the current user as a legitimate user or an impostor. Centeno et al. \cite{centeno2018mobile} applied a siamese convolutional neural network to learn smartphone users' behavioral patterns from raw sensor data. Based on the extracted deep features, they choose the one-class SVM as the classifier. Third, the authentication model’s weak capability of capturing smartphone users’ behavioral patterns from sensor data leads to unsatisfactory authentication performance \cite{shen2017performance, zhu2020hybrid, sitova2016hmog, yang2019personaia, li2018using, shen2022mmauth, wang2021framework, zhu2020espialcog}. Shen et al. \cite{shen2017performance} constructed 48 statistical features from each sensor to represent the smartphone user’s behavioral patterns and authenticate his identity. Experimental results show that they can achieve a false rejection rate of 5.03\%, and a false acceptance rate of 3.98\%. Zhu et al. \cite{zhu2020hybrid} proposed a hybrid deep learning method, which includes a convolutional neural network (CNN) architecture for mixture feature extraction, and a SVM classifier for effective model training. Experiments demonstrate that the proposed method can achieve 95.01\% authentication accuracy on a real-world dataset. 

To overcome the disadvantages of currently continuous authentication methods, we propose a relative attention-based one-class adversarial autoencoder architecture for continuous authentication of smartphone users. The proposed authentication architecture consists of four parts. First, we propose a one-class adversarial autoencoder to learn behavioral patterns of the legitimate user, which is trained only with the positive samples from the legitimate user in an unsupervised manner. Second, based on the latent representation from the encoder, a latent discriminator is applied to force latent representations of legitimate user's samples subject to spatial distribution uniformly. Third, a sample discriminator is trained to distinguish between the positive samples and the negative samples generated by the decoder with a prior distribution $\tilde z \in \mathbb{U}( - 1,{\text{ }}1)$. Note that the role of the sample discriminator is to allow the autoencoder to reconstruct higher-quality positive samples during the training phase, rather than to classify the access user as a legitimate user or an impostor during the test phase. Fourth, we modify the standard self-attention mechanism using convolution projection instead of linear projection to conduct the query, key and value maps. Compared with the standard self-attention mechanism \cite{vaswani2017attention}, the relative attention mechanism is more suitable for applications in the scenarios with limited computing power. Experimental results demonstrate that the effective stack of convolutional layer and the constructed relative attention layer can improve the model's capability of capturing contextual semantic information from behavioral biometrics. Besides, unlike recent approaches, training a one-class classifier to classify the access user as legitimate or an impostor, we apply a probabilistic method \cite{pidhorskyi2018generative} to compute the probability that the reconstructed sample is generated from the distribution of legitimate user's samples. To verify the feasibility of the proposed relative attention-based one-class adversarial autoencoder architecture, we design a continuous authentication system based on the proposed relative attention-based one-class adversarial autoencoder, which consists of four modules: sensor data acquisition module, data preprocessing module, relative attention-based one-class adversarial autoencoder module, and continuous authentication module. Experimental results show that the designed continuous authentication system can achieve excellent performance of 1.05\% EER, 1.09\% EER and 1.08\% EER on HMOG (Hand Movement, Orientation, and Grasp) dataset, BrainRun dataset and IDNet dataset, respectively.

In summary, the key contributions of the paper are outlined as follows:

\begin{itemize}
	\item 
	We propose a relative attention-based one-class adversarial autoencoder architecture to model the behavioral patterns of legitimate users, which solves the problem that negative samples from attackers are not easy to obtain in a real-world scenario.
	
	\item 
	We combine the constructed relative attention layers and convolutional layers to enhance the capability of capturing contextual semantic information from smartphone users' behavioral biometrics. 
	
	\item
	We design a continuous authentication system to evaluate the proposed relative attention-based one-class adversarial autoencoder architecture. Comprehensive evaluations and comparative experiments are performed to verify the effectiveness and the superiority of the proposed relative attention-based one-class adversarial autoencoder architecture on three public datasets.
	
\end{itemize}

The remainder of this article is organized as follows. Section \ref{related work} reviews the related work. We propose a relative attention-based one-class adversarial autoencoder architecture in Section \ref{autoencoder}. We detail the designed continuous authentication system in Section \ref{system}. Detailed performance evaluation experiments are presented in Section \ref{experiments}. Finally, we conclude this work in Section \ref{conclusion}.

\section{Related Work}
\label{related work}

\subsection{Behavioral Biometrics-based Continuous Authentication} 

Smartphones have a variety of built-in sensors, such as accelerometer, gyroscope, magnetometer, elevation, which can capture acceleration, angular velocity, orientation, and other information. Sensory data recorded by these sensors can measure the smartphone users’ behavioral patterns when interacting with smartphones.  

In recent years, behavioral biometrics-based continuous authentication has attracted the attention of many researchers, related researchers conducted the latest, comprehensive, extensive, and targeted investigations on continuous authentication based on behavioral biometrics \cite{stylios2021behavioral, abuhamad2020sensor}. Behavioral biometrics are categorized into the following categories: touch gesture-based authentication \cite{shen2016performance, alqarni2020identifying, yang2019behavesense, debard2018learning}, gait-based authentication \cite{giorgi2021using, shila2018adversarial, zou2020deep}, keystroke-based authentication \cite{krishnamoorthy2018identification, 2018Multi, zhang2016model, tharwat2018personal}, hand waving-based authentication \cite{sitova2016hmog, buriro2017please, yang2015unlocking}, and multi-model fusion \cite{li2018using, shen2017performance, li2021deffusion, abuhamad2020autosen, li2021cnn}, etc. Shen et al. \cite{shen2016performance} analyzed the feasibility and applicability of smartphone users’ touch-interaction behavior for continuous authentication on a real-world scenario dataset. Experimental results demonstrate that the touch behavior of smartphone user interacting with smartphones can well represent the user's unique behavioral patterns. Giorgi et al. \cite{giorgi2021using} proposed a continuous authentication architecture based on walking gait behavior analysis, which uses a recurrent neural network model for the authentication phase. In the experiments, different user walking behaviors or a combination of them are used to analyze the impact of walking type on smartphone user authentication. Unlike previous continuous authentication schemes that concentrated on the design and performance evaluation of deep learning models, Ray-Dowling et al. \cite{ray2022evaluating} focus more on the impact of different modalities of behavioral biometrics (such as swipe, hand movement, etc.) on the performance of continuous authentication. Additionally, they assessed the authentication performance using multiple feature modalities with both one-class SVM and binary SVM. Krishnamoorthy et al. \cite{krishnamoorthy2018identification} proposed a keystroke dynamics-based authentication scheme, which applies the support vector machine classifier to classify the access user as a legitimate user or an impostor. Besides, they employed minimum redundancy maximum relevance minimum redundancy maximum relevance (mRMR) feature selection method to improve authentication performance. Sitová et al. \cite{sitova2016hmog} introduced a set of behavioral features (HMOG) for continuous authentication of smartphone users. The HMOG features can capture subtle micro-movement and orientation dynamics resulting when smartphone users interact with smartphones. To solve the problem of insufficient training samples or uneven sample distribution in continuous authentication of smartphone users based on behavioral biometrics. Li et al. \cite{li2018using} applied five data augmentation approaches of permutation, sampling, scaling, cropping, and jittering to create additional data on the training samples. An effective feature fusion scheme plays an important role in improving the performance of continuous authentication. Li et al. \cite{li2021deffusion} proposed a lightweight convolutional neural network architecture to learn and extract valuable deep features from statistical features. Besides, they employed a balanced feature concatenation to fuse the extracted valuable features.

In this paper, we leverage sensory data from the accelerometer, gyroscope, and magnetometer to capture users’ behavioral patterns when users interact with smartphones.

\subsection{Self-attention Mechanism} 

The self-attention mechanism \cite{vaswani2017attention} has achieved great success in the field of natural language processing (NLP). Due to the effectiveness and scalability of self-attention mechanism, multiple works explore the combination of self-attention mechanism and convolutional neural networks in computer vision tasks.

Recently, pure transformer models or models combined pure transformer and convolutional neural networks are presented for computer vision tasks. Vision Transformer (ViT) \cite{dosovitskiy2020image} is the first to apply a standard transformer \cite{vaswani2017attention} to perform image classification tasks. When pretrained on a large-scale dataset, the performance of the ViT even surpass state-of-the-art ConvNets. To solve the problems faced by the migration of the standard transformer from the domain of natural language processing to the computer vision, Liu et al. \cite{liu2021swin} presented a hierarchical transformer, which applies shifted windows to compute the latent representation. Carion et al. \cite{carion2020end} proposed an encoder-decoder network for object detection (DETR), which applies the standard transformer to learn rich contextual semantic information. Inspired by DETR \cite{carion2020end}, Li et al. \cite{li2021medical} applied an effective squeeze-and expansion transformer layers for medical image segmentation. Jiang et al. \cite{jiang2021transgan} built a generative adversarial network architecture without convolution layers, which only uses the standard transformer. Han et al. \cite{han2021transformer} presented a transformer in transformer (TNT) network for image recognition. The TNT applies an outer transformer to learn global dependencies, and employs an inner transformer to extract useful features from pixel level.

Although the pure transformer has been proven to improve the performance of traditional convolutional neural networks (CNN), it also brings huge computational overhead, especially at high-resolution input. Many works employ self-attention within limited region (e.g., 5×5 grid). Wang et al. \cite{wang2021evolving} combined the self-attention mechanism and residual learning with training a very deep residual network. Ramachandran et al. \cite{ramachandran2019stand} applied a stand alone self-attention block to replace the core building block of ResNet \cite{he2016deep}. Wu et al. \cite{wu2021cvt} applied a convolutional projection to replace the linear projection in the pure transformer module. Li et al. \cite{li2022contextual} proposed a contextual transformer block to strength the capability of learning rich contextual semantic information among neighbor keys. 

Considering that the proposed continuous authentication system is deployed on smartphones, this article presents a relative attention layer, which applies convolutional layers to modify the standard self-attention mechanism \cite{vaswani2017attention}. Unlike some previous works that applied self-attention block to replace the convolutionla layers or as an enhancement on top of the convolutions, this article learns richer contextual information from behavioral biometrics through the effective stacking of relative attention layers and convolutional layers.

\section{Architecture of Relative Attention-based One-class Adversarial Autoencoder}
\label{autoencoder}

\begin{figure*}[!t]
	\centering
	
	\includegraphics[width=1.0\linewidth]{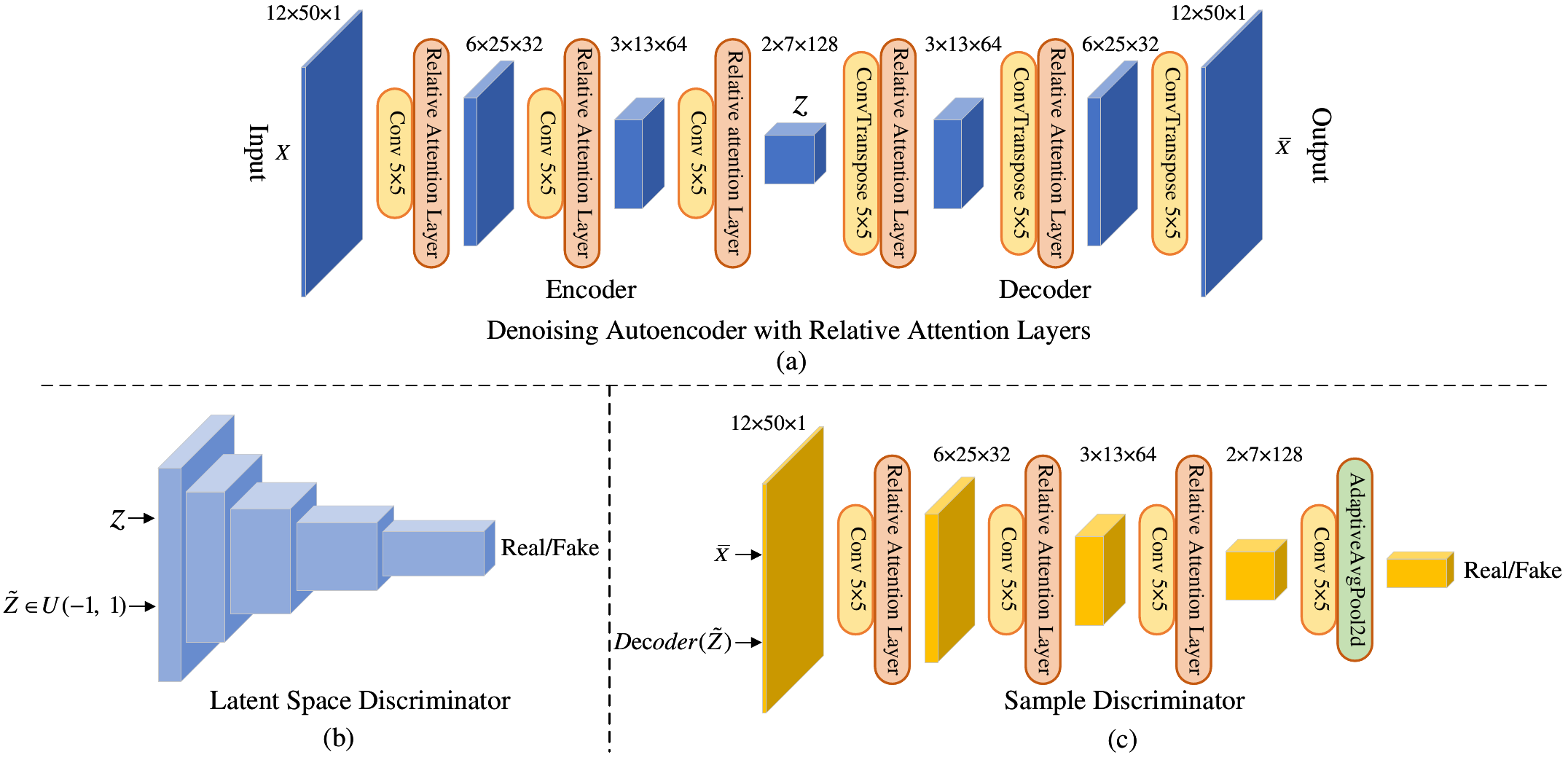}
	
	\caption{The architecture of relative attention-based one-class adversarial autoencoder: the denoising autoencoder with relative attention layer, the latent space discriminator, and the sample discriminator with relative attention layer.}
	\label{figure1}
\end{figure*}

As shown in Fig. \ref{figure1}, the proposed relative attention-based one-class adversarial autoencoder consists of four components: the relative attention layer, the denoising autoencoder, the latent space discriminator, and the sample discriminator. We describe each component below.

\subsection{Relative Attention Layer}

\begin{figure}[!t]
	\centering
	
	\includegraphics[width=0.90\linewidth]{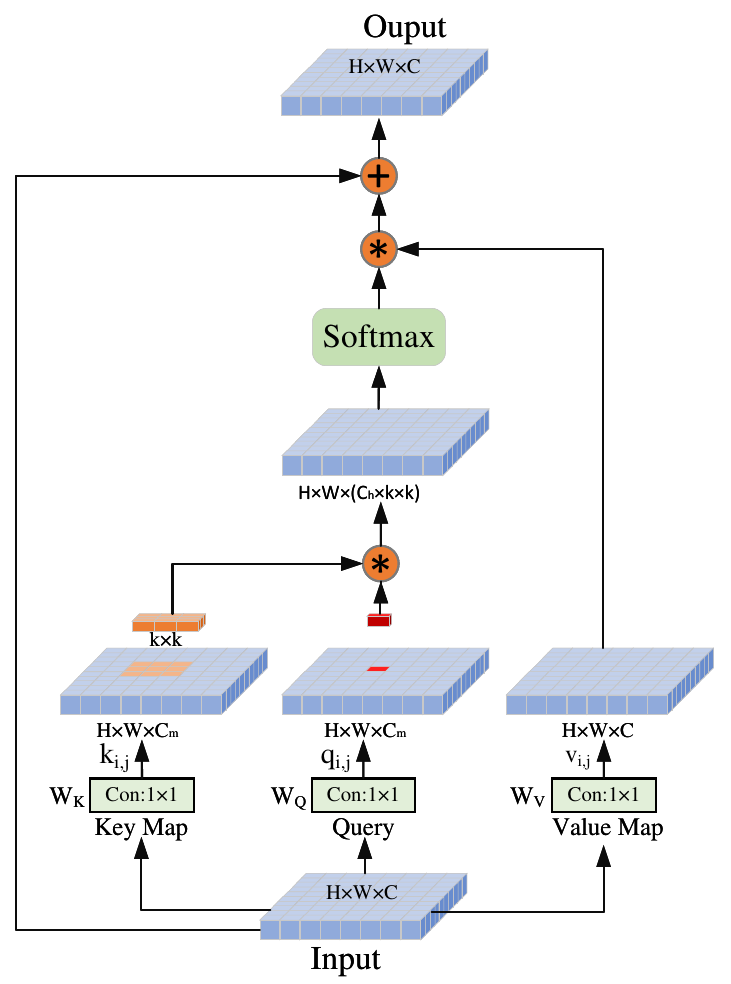}
	
	\caption{The relative attention layer, learning the dependence between a unit and its neighborhoods.}
	\label{figure2}
\end{figure}

The original self-attention mechanism \cite{vaswani2017attention} aims to learn the dependency between a unit and the entire input sequence. The attention value for a unit can be computed as:
\begin{equation}
	\label{eq:attweight}
	\centering
	{\textrm{AttWeight}}(x,x) = f(K(x),Q(x))
\end{equation}
\begin{equation}
	\label{eq:attention value}
	\centering
	{\textrm{Attention}}(x) = {\textrm{AttWeight}}(x,x) \cdot V(x)
\end{equation}
\begin{equation}
	\label{eq:convolution1}
	\centering
	{x_{{\textrm{out}}}} = {\textrm{MLP}}({\textrm{Attention}}(x))
\end{equation}
where the ${\textrm{AttWeight}}(x,x)$ is the attention matrix between each unit from the input sequence and the entire input sequence. The Q, K, V represent the queries, keys, values mapping, respectively, and the $f$ is the softmax function. The \textbf{query} represents an input element and is used to score each \textbf{key}. Derived from the input data, the query vector determines the level of attention (or weight) that each part of the input should receive. The \textbf{key} is associated with each input element and is used for matching with a \textbf{query}. When a \textbf{query} is compared with a \textbf{key}, the resulting score represents the relevance or attention that the corresponding \textbf{value} should receive. Once the scores between \textbf{query} and \textbf{key} are computed, they are used to weight the \textbf{value}. The weighted values are then summed up to produce the final output for each input element, which is a representation that has been informed by the other elements in the input according to their computed relevance. As shown in equation (\ref{eq:attweight}) and (\ref{eq:attention value}), the $Q$ and $K$ are used to compute the similarity between each query and corresponding keys. The ${x_{{\textrm{out}}}}$ is the final output with multi-layer perceptron (MLP) transformation.

Taking into account the computational overhead of the original self-attention mechanism \cite{vaswani2017attention}. In this article, we apply convolutional layers to perform the query, key and value embeddings on 2D features map respectively, which is light-weight
enough for mobile platform. For a 2D feature map, give a unit ${x_{i,j}}$, we compute the relative attention weight between ${x_{i,j}}$ and the neighborhood ${x_{a,b}}$ ($a,b \in \eta (i,j)$), where the $\eta (i,j)$ is a set of neighbors in a fixed domain with ${x_{i,j}}$ as the center, e.g., a k×k grid centered at $(i,j)$ in a 2D feature map. As shown in Fig. \ref{figure2}, we apply convolutional projections to perform the query, key, value embeddings, respectively. Then the relative attention value for the ${x_{i,j}}$ is computed as follows:
\begin{equation}
	\label{eq:convolution29}
	\centering
	{y_{i,j}} = \sum\limits_{a,b \in \eta (i,j)} {\frac{{\exp (q_{i,j}^T{k_{a,b}})}}{{\sum\limits_{m,n \in \eta (i,j)} {\exp (q_{i,j}^T{k_{m,n}})} }}{v_{a,b}}}
\end{equation}

\begin{equation}
	\label{eq:convolution9}
	\centering
	\begin{split}
		{q_{i,j}} = {W_Q}{x_{i,j}}\\
		{k_{i,j}} = {W_K}{x_{i,j}}\\
		{v_{i,j}} = {W_V}{x_{i,j}}
	\end{split}
\end{equation}
where the ${y_{i,j}} \in {\mathbb{R}^{{d_{out}}}}$ is the output at ${x_{i,j}}$, and ${q_{i,j}}$, ${k_{i,j}}$, ${v_{i,j}} \in {\mathbb{R}^{{d_{out}}}}$ are the intermediate value produced by ${x_{i,j}}$ and its neighborhoods $\eta (i,j)$. The ${q_{i,j}}$ and ${k_{i,j}}$ compute the similarity between each query and corresponding keys with the local $k \times k$. The choice of k in a fixed k×k neighborhood is crucial for balancing contextual capture and computational efficiency. In this paper, we employ Neural Architecture Search (NAS) \cite{zoph2016neural} techniques to determine the optimal size of the convolutional kernel k. NAS allows us to automatically explore a wide range of architectural configurations and identify the most effective k that maximizes model performance while considering computational constraints. This approach enables the model to dynamically adapt to varying behavioral biometrics characteristics and capture both local and long-range dependencies effectively. The embedding matrices ${W_Q}$, ${W_K}$, ${W_V} \in {\mathbb{R}^{{d_{out}} \times {d_{in}}}}$ are implemented with $1\times1$ convolutional layers.

\subsection{Denoising Autoencoder}
\label{sssec:stem}

We employ a denoising autoencoder \cite{vincent2008extracting} network to learn latent representation from input. Compared with the standard autoencoder, the denoising autoencoder add noise to the input, and attempts to reconstruct the input. The denoising autoencoder can reduce overfitting and make the trained encoder more robust, thereby enhancing the generalization ability of the denoising autoencoder. Besides, as shown in Fig. \ref{figure1}a, unlike the prior denoising autoencoder architecture, we improve the representation learning capability and capacity of the model through the effective stacking of convolutional layers and relative attention layers. We expect the designed denoising autoencoder to be able to reconstructed samples as if they are drawn from the real distribution of the input.  

The reconstruction loss for the denoising autoencoder can be defined as:
\begin{equation}
	\label{reconstruction loss}
	\centering
	{L_{rec}} = \parallel x - f(g(x+n))\parallel ^2
\end{equation}
where $x$ is the input, $f$ is the Decoder, $g$ is the Encoder, and $n \sim \mathcal{N}(0,0.2)$.

\subsection{Latent Space Discriminator}

In order for the encoder to encode the inner class samples to the latent space representation $z$ with distribution close to the prior distribution $\tilde z  \in \mathbb{U}( - 1,{\text{ }}1)$. As shown in Fig. \ref{figure1}b, we apply a latent space discriminator to force latent space representations of inlier class samples to be distributed uniformly across the prior distribution. The latent space discriminator is applied to distinguish between the low-dimensional latent space representations of inlier class samples and the samples drawn from the prior distribution $\tilde z$. The adversarial loss of the latent space discriminator $D_l$ can be formulated as:

\begin{equation}
	\label{eq:convolution31}
	\centering
	{L_{latent}} = E\left[ {\log {D_l}(\tilde z)} \right] + E\left[ {\log (1 - {D_l}(g(x+n)))} \right]
\end{equation}
The weights of $g$ are updated to minimize this objective and the $D_l$ tries to maximize it. The ultimate goal is to expect the low-dimensional latent space representation of inlier class samples to be distributed following the prior distribution $\tilde z$.

\subsection{Sample Discriminator}

A sample discriminator ${D_s}$ is trained to distinguish between the positive samples and the negative samples generated from the decoder. The positive samples $x$ are from the inlier class, and the negative samples $D{\textrm{e}}coder\left( {\tilde z} \right)$ are generated by the decoder from the latent space with a prior distribution $\tilde z \in \mathbb{U}( - 1,{\text{ }}1)$. As shown in Fig. \ref{figure1}c, the sample discriminator architecture is composed of convolutional layers and relative attention layers. The decoder attempts to generate the negative samples that can fool the sample discriminator, and the sample discriminator learns to distinguish between the positive samples and the negative samples. The autoencoder and sample discriminator are trained as an adversarial game. We leverage the adversarial loss to improve the quality of reconstructed samples by autoencoder. The adversarial loss for the sample discriminator is formulated as:

\begin{equation}
	\label{sample discriminator}
	\centering
	{L_{sample}} = E\left[ {\log \left( {{D_s}\left( x \right)} \right)} \right] + E\left[ {\log \left( {1 - {D_s}\left( {Decoder(\tilde z)} \right)} \right)} \right]
\end{equation} 

\begin{figure*}[!t]
	\centering
	
	\includegraphics[width=0.95\linewidth]{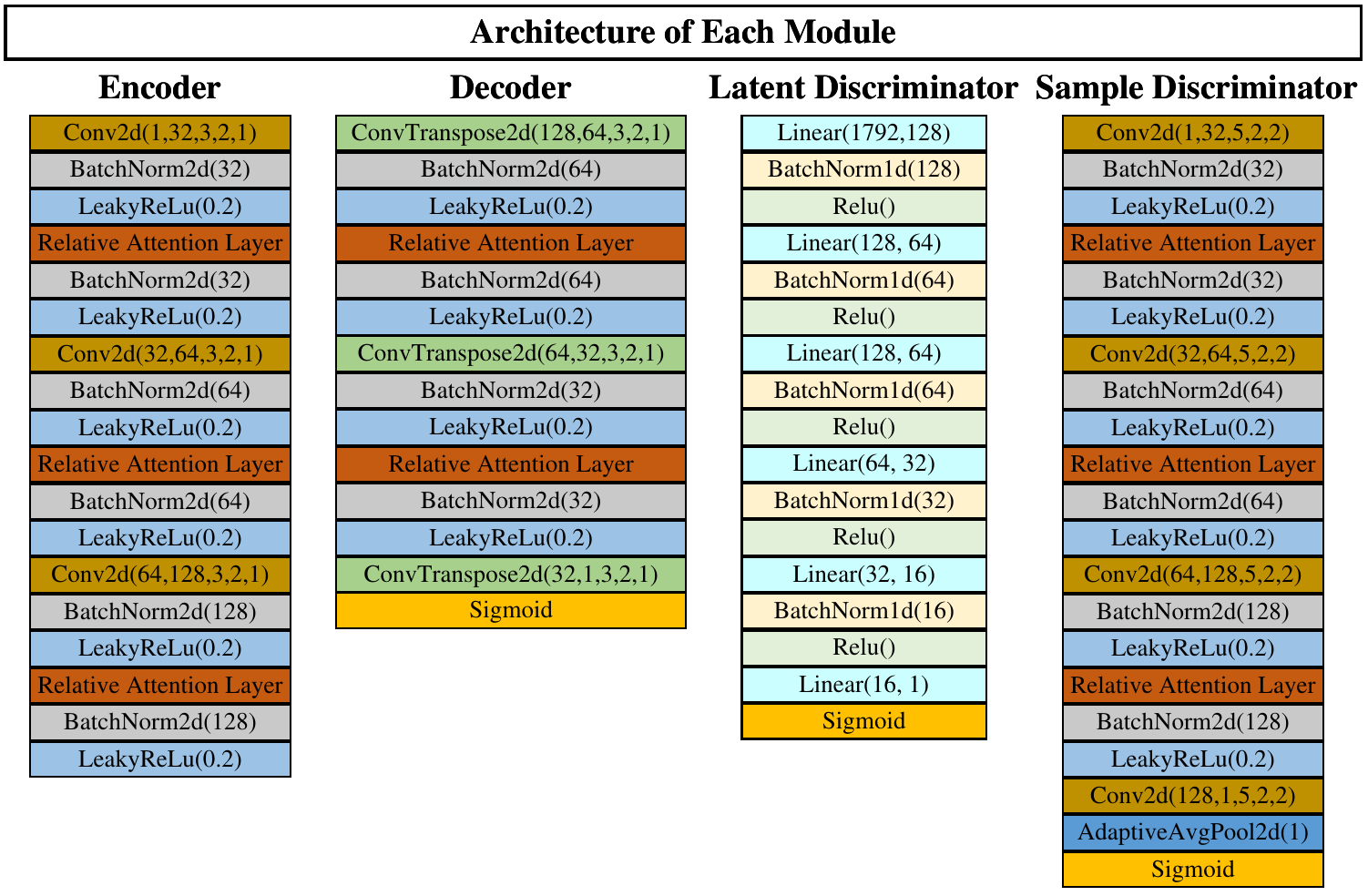}
	
	\caption{Architecture of the relative attention-based one-class adversarial autoencoder. Different layers are represented with different colors.}
	\label{figure6}
\end{figure*} 

\begin{algorithm}[!t]
	\caption{Training methodology of the relative attention-based one-class adversarial autoencoder. ${D_l}$ and $D_s$ represent the latent discriminator and the sample discriminator, respectively. En and De are the encoder and the decoder, respectively.}
	\label{alg1}
	\KwIn{Training set $x$, number of iterations $N$, }
	\KwOut{En, De, $D_l$, $D_s$}
	\For{iteration=$1$ \KwTo $\rightarrow$ $N$}{
		$n \leftarrow \mathcal{N}(0,0.2)$\\
		$z$  $\leftarrow$ En($x$ + n)\\
		$\tilde z$  $\leftarrow$  $\mathbb{U}( - 1,1)$\\
		Sample discriminator update:\\
		${L_{\text{sample}}} \leftarrow {D_s}(De(\tilde z),0) + {D_s}(x,1)$\\
		Back-propagate ${L_{\text{sample}}}$ to update $D_s$\\
		Latent discriminator update:\\
		${L_{latent}} \leftarrow {D_l}(z,0) + {D_l}(\tilde z,1)$\\
		Back-propagate ${L_{latent}}$ to update ${D_l}$\\
		~\\
		Autoencoder update:\\
		${L_{rec}} \leftarrow \parallel x - De(z)\parallel ^2$\\
		${L_{\text{sample}}} \leftarrow {D_s}(De(\tilde z),1) + {D_s}(x,0)$\\
		${L_{latent}} \leftarrow {D_l}(z,1) + {D_l}(\tilde z,0)$\\
		Back-propagate $\lambda {{\text{L}}_{rec}}$ + ${L_{sample}}$ + ${L_{latent}}$ to update $En$, $De$\\
		
	}
\end{algorithm}

\subsection{Implementation Details}

Detailed network architecture of the relative attention-based one-class adversarial autoencoder is shown in Fig. \ref{figure6}. The encoder contains three convolutional layer and three relative attention layers. For each convolutional layer and relative attention layer, followed by the batch normalization and the leaky Relu operations. The decoder consists of three transposed convolutional layers and two relative attention layers. These are, then followed by the batch normalization and the leaky Relu operations, except for the last transposed convolutional layer. A sigmoid activation operation is applied after the last transposed convolutional layer. The latent discriminator contains six linear layers, and each linear layer is followed by the batch normalization and the Relu operations, except for the last linear layer. And, a sigmoid activation function is placed after the last linear layer. The sample discriminator contains four convolutional layers and three relative attention layers followed by the batch normalization and the leaky Relu operations, except for the last convolutional layer. An adaptive average pooling layer and a sigmoid activation function are placed after the last convolutional layer. The training step of the relative attention-based one-class adversarial autoencoder is summarized as Algorithm \ref{alg1}, and the $\lambda$ is set to 10.

During the training phase, we implemented several strategies to enhance stability and convergence. First, we applied spectral normalization to the discriminators, which helps in constraining the Lipschitz constant of the network and mitigates training instability. We also incorporated gradient penalty techniques to further stabilize the adversarial training process by enforcing smoothness in the discriminator's output with respect to its input. Additionally, we adopted a balanced training regimen by carefully tuning the learning rates and employing batch normalization, which contributed to minimizing the risk of mode collapse and improving overall training dynamics.

\section{Continuous Authentication System}
\label{system}

\begin{figure*}[!t]
	\centering
	
	\includegraphics[width=1.0\linewidth]{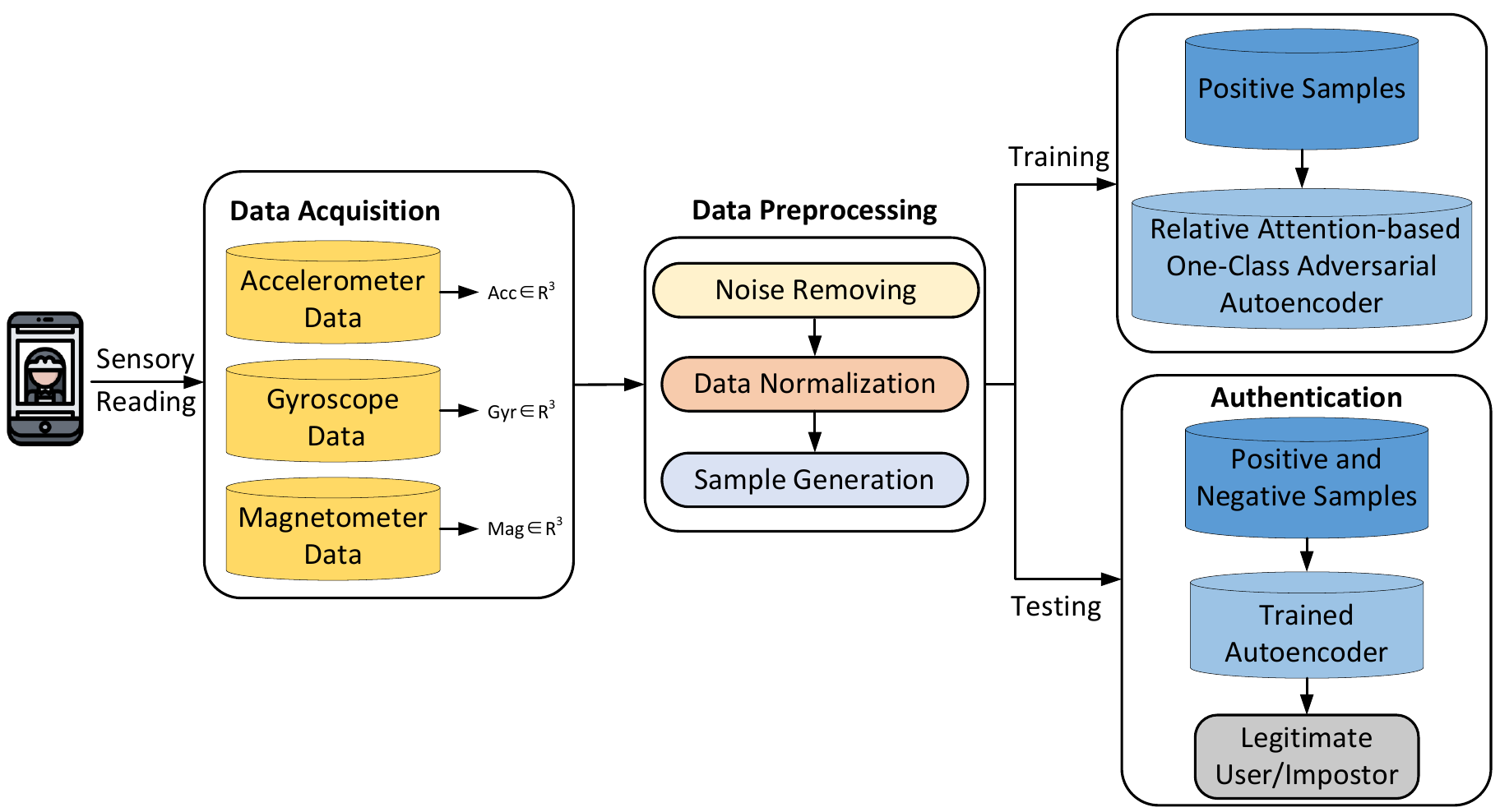}
	
	\caption{Architecture of the proposed continuous authentication system.}
	\label{figure4}
\end{figure*}

To evaluate the proposed relative attention-based one class adversarial autoencoder, we design and implement a comprehensive continuous authentication system. As shown in Fig. \ref{figure4}, the proposed continuous authentication system is composed of three modules: 1) data acquisition module; 2) data preprocessing module; 3) continuous authentication module. We will describe the implementation of each module in detail.

\begin{figure*}[!t]
	\centering
	
	\includegraphics[width=1.0\linewidth]{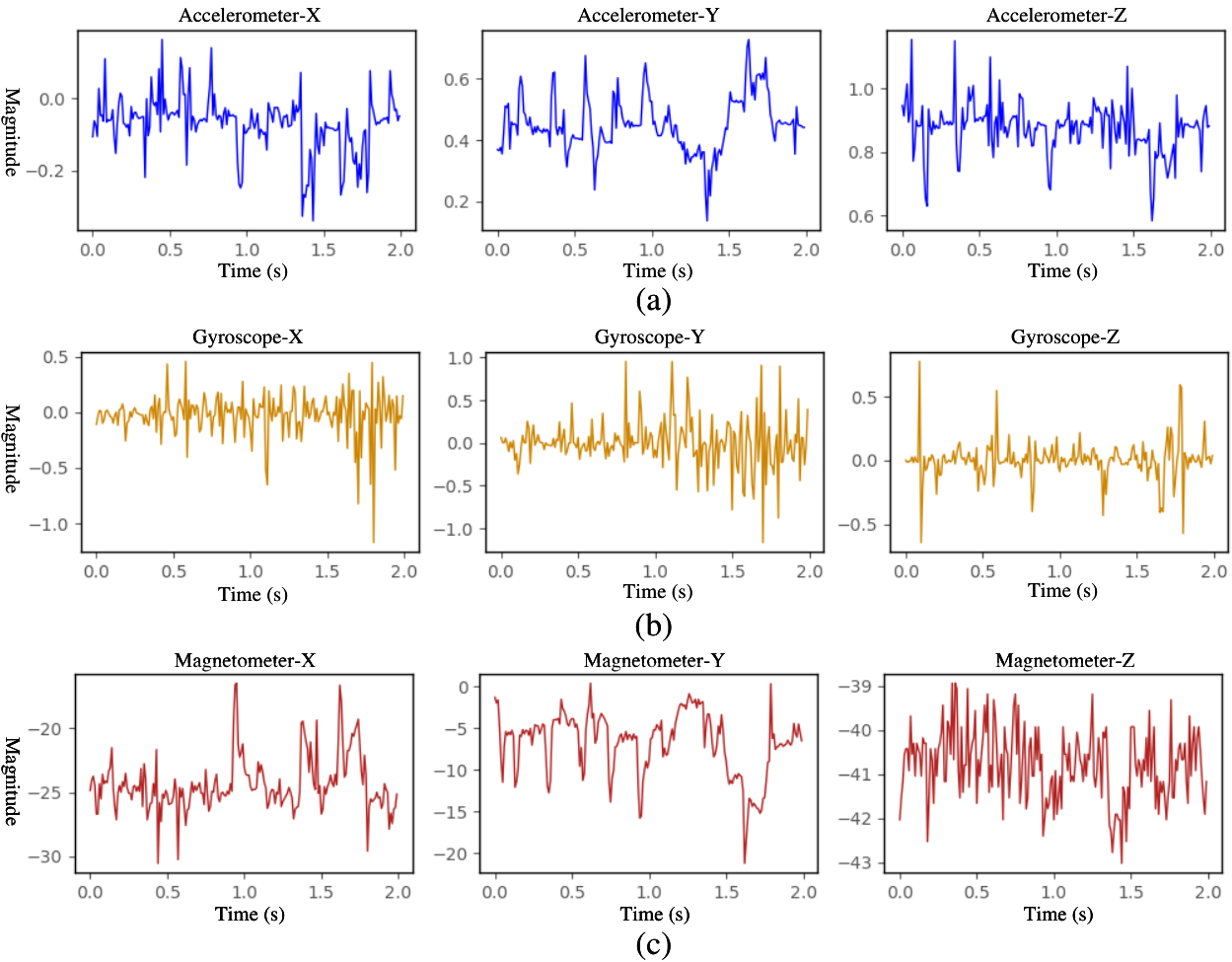}
	
	\caption{The sensory readings of three-axis from different sensors. (a) sensory data sequence of three-axis from the accelerometer; (b) sensory data sequence of three-axis from the gyroscope; (c) sensory data sequence of three-axis from the magnetometer.}
	\label{figure5}
\end{figure*}

\subsection{Data Acquisition}

When the user interacts with the smartphone, the sensory data collected by the smartphone's built-in sensors can imply the user's behavioral biometrics. The proposed continuous authentication system applies data collected from three sensors (accelerometer, gyroscope, magnetometer) to model smartphone user's behavioral patterns and then authenticate smartphone user's identity. The accelerometer is applied to measure the magnitude and direction of the acceleration of the smartphone due to force. It can record the movement patterns of smartphone user when interacting with smartphone. The gyroscope can measure the rotation rate of the smartphone when user interacts with smartphone. The magnetometer can measure the strength of the surrounding magnetic field and locate the location of the smartphone. The proposed continuous authentication system can learn users' behavioral patterns well by combining sensory data from three sensors.
\subsection{Data Preprocessing}

To enable the relative attention-based one-class adversarial autoencoder to learn rich contextual semantic information that can represent the legitimate user's behavioral patterns, we perform noise removing, data normalization, sample generation on the raw behavioral biometrics.

\textbf{Noise Removing:} The noise contained in sensor data reading make a significant impact on the authentication performance, which can be generated from sensor data collection stage. These noises may be generated by participants' irregular operations during sensor data collection. Therefore, noise should be removed to improve the sensor data readings' ability of characterizing smartphone user's behavioral patterns. There are two types of outliers that we will treat as noise: 1) signal mutation in smooth sensor signal, noise can cause some peaks in the smoothed sensor signal; 2) unchanged sensor signal for a period of time. During sensor data collection, participants may not interact with  smartphones for a period of time, for example, putting the smartphone on the desk.

\textbf{Data Normalization:} As shown in Fig. \ref{figure5}, the biometric data collected by different sensors are significant difference in numerical values. If the biometric data are directly used as the input of the proposed relative attention-based one-class adversarial autoencoder, the deep learning model may highlight the role of larger values of biometric data, while relatively weakening the role of smaller values of biometric data. However, those biometric data with smaller values may play a decisive role in the final prediction results. Besides, the biometric data collected by multiple sensors may contain some noise and outliers. To reduce the large difference in biometric data distribution and the influence of noise and outliers on the authentication performance, this article normalizes the biometric data from multiple sensors before fed to the relative attention-based one-class adversarial autoencoder.

In this article, min-max normalization is applied to perform a linear transformation on the raw biometric data. The biometric data sequence for each sensor in one axis can be represented as $({x_1},{x_2},{x_3},{x_4}, \cdot  \cdot  \cdot ,{x_{n - 1}},{x_n}) \in {\mathbb{R}^n}$, then we perform the normalization by:

\begin{equation}
	\label{eq:convolution25}
	\centering
	{y_i} = \frac{{{x_i} - \mathop {\min }\limits_{1 \le j \le n} \{ {x_j}\} }}{{\mathop {\max }\limits_{1 \le j \le n} \{ {x_j}\}  - \mathop {\min }\limits_{1 \le j \le n} \{ {x_j}\} }}
\end{equation}
where the transformed data for each sensor in one axis can be represented as $({y_1},{y_2},{y_3},{y_4}, \cdot  \cdot  \cdot ,{y_{n - 1}},{y_n}) \in [0,1]$.

\textbf{Sample Generation:} Based on raw biometric data collected from multiple sensors (accelerometer, gyroscope, magnetometer), we need to process them into training samples for the input of the proposed relative attention-based one-class adversarial autoencoder. For each time point, the collected behavioral biometric data can be represented as $s = (x,y,z) \in {\mathbb{R}^3}$, where $x$, $y$, $z$ is the three-axis of the sensory data. Besides, we also add a fourth axis for each sensor at each time point, called Magnitude (M). The magnitude can be formulated as 
$M = \sqrt {{x^2} + {y^2} + {z^2}}$. Then for each time point, the behavioral biometric data from three sensors can be represented as $({x_{acc}},{y_{acc}},{z_{acc}},{m_{acc}},{x_{gyr}},{y_{gyr}},{z_{gyr}},{m_{gyr}},{x_{mag}},{y_{mag}},\\{z_{mag}},{m_{mag}}) \in {\mathbb{R}^{12}}$, where the acc, gyr, mag indicate the accelerometer, gyroscope, magnetometer, respectively. We apply a time window to segment the biometric data readings from three sensor without time overlap. Then for a time window $t$, the biometric data reading from three sensors can be represented as a $d \times n$ matrix:  

\begin{equation}
	\label{eq:convolution27}
	\centering
	S = \left[ {\begin{array}{*{20}{c}}
			{x_{acc}^1}&{x_{acc}^2}& \cdots &{x_{acc}^{n - 1}}&{x_{acc}^n} \\ 
			{y_{acc}^1}&{y_{acc}^2}& \cdots &{y_{acc}^{n - 1}}&{y_{acc}^n} \\ 
			{z_{acc}^1}&{z_{acc}^2}& \cdots &{z_{acc}^{n - 1}}&{z_{acc}^n} \\ 
			{m_{acc}^1}&{m_{acc}^2}& \cdots &{m_{acc}^{n - 1}}&{m_{acc}^n} \\ 
			{x_{gyr}^1}&{x_{gyr}^2}& \cdots &{x_{gyr}^{n - 1}}&{x_{gyr}^n} \\ 
			{y_{gyr}^1}&{y_{gyr}^2}& \cdots &{y_{gyr}^{n - 1}}&{y_{gyr}^n} \\ 
			{z_{gyr}^1}&{z_{gyr}^2}& \cdots &{z_{gyr}^{n - 1}}&{z_{gyr}^n} \\ 
			{m_{gyr}^1}&{m_{gyr}^2}& \cdots &{m_{gyr}^{n - 1}}&{m_{gyr}^n} \\ 
			{x_{mag}^1}&{x_{mag}^2}& \cdots &{x_{mag}^{n - 1}}&{x_{mag}^n} \\ 
			{y_{mag}^1}&{y_{mag}^2}& \cdots &{y_{mag}^{n - 1}}&{y_{mag}^n} \\ 
			{z_{mag}^1}&{z_{mag}^2}& \cdots &{z_{mag}^{n - 1}}&{z_{mag}^n} \\ 
			{m_{mag}^1}&{m_{mag}^2}& \cdots &{m_{mag}^{n - 1}}&{m_{mag}^n} 
	\end{array}} \right]
\end{equation}
where the d = 12, and the $n = t*f$. We resample the biometric data readings with the sample rate $f = 100Hz$, and set the time window $t=0.5$s. Then the training samples have a shape of $12$×$50$, which are fed to the relative attention-based one-class adversarial autoencoder for training.

\subsection{Continuous Authentication Model}

Unlike the current continuous authentication methods based on deep learning, which require impostors' behavioral biometric data (negative samples) to train the deep learning-based continuous authentication model. This article proposes a relative attention-based one-class adversarial autoencoder to continuously authenticate the smartphone user's identity, which does not require negative samples to train the model throughout the training phase. Therefore, the proposed continuous authentication model will not cause the leakage of smartphone users' behavioral biometric data. We have described the composition of relative attention-based one-class adversarial autoencoder in detail in Section \ref{autoencoder}.

In the authentication phase, we apply a probabilistic-based prediction approach to evaluate the reconstructed sample how likely it is generated by the distribution of the legitimate smartphone user's samples. Then we predict whether the test sample comes from a legitimate user or an impostor based on the calculated probability value. Give a sample $x \in {\mathbb{R}^m}$, it can be modeled as:

\begin{equation}
	\label{eq:convolution24}
	\centering
	\begin{gathered}
		x = x^\parallel + \varepsilon  \hfill = f(\tilde z) + \varepsilon  \hfill = f(g(x)) + \varepsilon  \hfill \\ 
	\end{gathered}
\end{equation}
where $f$ is the decoder, $g$ is the encoder. $x$ can be non-linearly projected into ${x^\parallel } \in \mathcal{M}$. And ${x^\parallel } = f(\tilde z)$ is low-dimensional manifold $\tilde z \in {\mathbb{R}^n}$ embedded to high-dimensional space with $m > n$, where $\tilde z = g(x)$. For the low-dimensional manifold $\tilde z$, linearization based on the first order Taylor expansion can be formulated as:

\begin{equation}
	\label{eq:convolution14}
	\centering
	\begin{gathered}
		f(z) = f(\tilde z) + {J_f}(\tilde z)(z - \tilde z) + o(\parallel z - \tilde z{\parallel ^2}) \hfill \\ 
	\end{gathered}
\end{equation}
where $J_f$ is the Jacobi matrix at point $\tilde z$. Defining $\Gamma $ is the tangent space of $f$ at $\tilde x$, the $\Gamma $ is formulated as:

\begin{equation}
	\label{eq:convolution32}
	\centering
	\Gamma  = span({J_f}(\tilde z)) = span({U^\parallel }S{V^{\textrm T}}) = span({U^\parallel })
\end{equation}
where ${U^\parallel }S{V^{\textrm T}}$ is SVD decomposition of the Jacobi matrix, and $U$ is a unitary matrix. Then the sample $x$ can be decomposed into two parts with the tangent space and the space orthogonal to it:

\begin{equation}
	\label{eq:convolution26}
	\centering
	y = {U^{\textrm T}}x = \left[ {\begin{array}{*{20}{c}}
			{{U^{{\parallel ^{\textrm T}}}}x} \\ 
			{{U^{{ \bot ^{\textrm T}}}}x} 
	\end{array}} \right] = \left[ {\begin{array}{*{20}{c}}
			{{y^\parallel }} \\ 
			{{y^ \bot }} 
	\end{array}} \right]
\end{equation}
where ${y^\parallel }$ is parallel to $\Gamma $, and ${y^ \bot }$ is orthogonal to $\Gamma $, which is defined as a noise to make the sample $x$ away from the manifold distribution $\mathcal{M}$. Then given a sample $x$, its probability prediction function ${p_x}(x)$ can be formulated as:

\begin{equation}
	\label{eq:convolution36}
	\centering
	{p_x}(x) = {p_y}({U^{\textrm T}}x) = {p_y}({y^\parallel },{y^ \bot }){\kern 1pt}  = {p_{{y^\parallel }}}({y^\parallel }) * {p_{{y^ \bot }}}({y^ \bot })
\end{equation}

Given a test sample $x \in {\mathbb{R}^m}$ and its non-linear projection ${x^\parallel } \in \mathcal{M} \subset {\mathbb{R}^m}$. It is assumed that $x \approx f(g(x))$, then the ${p_{{y^\parallel }}}({y^\parallel })$ can be formulated as:

\begin{equation}
	\label{eq:convolution37}
	\centering
	\begin{gathered}
		{p_{{y^\parallel }}}({y^\parallel }) = {p_{{y^\parallel }}}({U^{{\parallel ^{\textrm T}}}}x) = {p_{{y^\parallel }}}({U^{{\parallel ^{\textrm T}}}}(x - {x^\parallel }) + {U^{{\parallel ^{\textrm T}}}}{x^\parallel }) \hfill \\
		{\kern 32pt} = {p_{{y^\parallel }}}({U^{{\parallel ^{\textrm T}}}}(x - f(g(x))) + {U^{{\parallel ^{\textrm T}}}}{x^\parallel }) \approx {p_{{y^\parallel }}}({U^{{\parallel ^{\textrm T}}}}{x^\parallel }) \hfill \\
		{\kern 32pt} = {p_{{x^\parallel }}}({x^\parallel }) = {p_{{x^\parallel }}}(f(z)) = \det {(US{V^{\textrm T}})^{ - 1}} * {p_z}(z) \hfill \\
		{\kern 32pt} = \det {S^{ - 1}} * {p_z}(z) \hfill \\ 
	\end{gathered}
\end{equation}

For each sample $x$, assuming that the noise is randomly distributed on the orthogonal manifold distribution. Then the intensity of noise ${y^ \bot }$ can be approximated by its distance to the center point of hypersphere ${S^{m - n - 1}}$. Then the ${p_{{y^ \bot }}}({y^ \bot })$ can be formulated as:

\begin{equation}
	\label{eq:convolution38}
	\centering
	{p_{{y^ \bot }}}({y^ \bot }) \approx \frac{{\Gamma (\frac{{m - n}}{2})}}{{2{\pi ^{\frac{{m - n}}{2}}}\parallel {y^ \bot }{\parallel ^{m - n}}}}{p_{\parallel {y^ \bot }\parallel }}(\parallel {y^ \bot }\parallel )
\end{equation}
where the $\Gamma ( \cdot )$ is the gamma function. The ${p_{\parallel {y^ \bot }\parallel }}(\parallel {y^ \bot }\parallel )$ can be learned offline by calculating the reconstruction error of the legitimate smartphone user's samples.

Given a sample $x$, if ${p_x}(x) \geqslant \tau$, the sample $x$ is from the legitimate user, otherwise the sample $x$ is from the impostor. The $\tau$ is a predefined threshold. 

To address the challenge of evolving and changing user behavior over time, we have implemented two measures: 1) During the training phase, we employ a substantial and diverse array of behavioral biometric data, aiming to capture as many common interaction patterns with smartphones as possible; 2) Our continuous authentication architecture includes a mechanism for collecting new user behavioral biometric data at regular intervals. This allows us to continually update and retrain the existing one-class model, ensuring it adapts to and learns new behavior patterns without losing the information it has previously acquired.

\section{Experiments}
\label{experiments}

Comprehensive experiments are performed to verify the effectiveness and superiority of the proposed continuous authentication system based on a relative attention-based one-class adversarial autoencoder. The network architecture is implemented with the PyTorch library, and we train the continuous authentication model on a Tesla T4 GPU for 100 epochs. The learning rate for the encoder, the decoder, the latent discriminator, and the sample discriminator are 0.00005, 0.0003, 0.00001, and 0.0001, respectively. The batch size is set to 16. Only the samples of legitimate user (positive samples) are needed to train the proposed relative attention-based one-class adversarial autoencoder during the training phase. We use 10 fold-cross-validation to train the continuous authentication model, and report the mean experimental results. We chose the threshold $\tau$ that provides the best trade-off between security and usability, which was the point where the True Positive Rate (TPR) first exceeded 0.97. And then we compute the FAR, FRR, and EER with the threshold $\tau$ in the experiments. After conducting numerous experiments, we have set the threshold $\tau$ for three different datasets (HMOG, BrainRun, IDNet) to be 4.022, 4.981 and 5.536, respectively. The main goals of the evaluation experiments are as follows: 1) performance on different public datasets; 2) comparison with representative continuous authentication methods for the continuous authentication task; 3) the effectiveness of the relative attention layer in proposed one-class adversarial autoencoder architecture; 4) performance with different time windows; 5) robustness against random attacks; 6) overhead analysis.

\subsection{Evaluation Metrics}

Four evaluation metrics are used to evaluate the performance of the proposed continuous authentication system. The AUROC is the area under the Receiver Operating Characteristics (ROC) curve. The False Acceptance Rate (FAR) refers to the proportion of the number of times that illegitimate users are incorrectly authenticated as legitimate users to the total number of times that should be authenticated as the impostors, which can be formulated as $FAR = \frac{{{\text{FP}}}}{{{\text{FP + TN}}}}$. The False Rejection Rate (FRR) refers to the proportion of the number of times that legitimate users are incorrectly authenticated as the impostors to the total number of times that should be authenticated as legitimate users, formulating as $FRR = \frac{{{\text{FN}}}}{{{\text{FN + TP}}}}$. Equal Error Rate (EER) is defined as the point where the FAR equals the FRR. True Positive (TP) indicates the number of normal samples are correctly predicted as normal samples. False Positive (FP) indicates the number of anomalous samples are incorrectly predicted as normal samples. True Negative (TN) indicates the number of anomalous samples correctly are predicted as anomalous samples. False Negative (FN) indicates the number of normal samples are incorrectly predicted as anomalous samples.

\subsection{Dataset}

We evaluate the performance of the proposed continuous authentication system on three public dataset, which can be used for continuous authentication of smartphone users.

\textbf{HMOG dataset:} HMOG \cite{sitova2016hmog} is a new set of behavioral biometric data for continuous authentication of smartphone users, which contains the Hand Movement, Orientation, and Grasp (HMOG). The behavioral biometric data are collected from the embedded sensors (accelerometer, gyroscope, and magnetometer), when the user interacts with the smartphone. The dataset contains the behavioral biometric data of 100 participants (53 male and 47 female). For each participant, they collected an average of 1193 taps for each session and 1019 key presses. The average duration of collecting biometric data for each session is 11.6 minutes.

\textbf{BrainRun dataset:} BrainRun \cite{papamichail2019brainrun} applies a built-in gesture capture tool to capture different types of gestures in the sliding behavior of the user when interacting with the smartphone. The BrainRun dataset mainly contains three parts of behavioral biometric data. The gesture data contains the coordinate information of the screen points generated by each participant's tapping and swiping when interacting with the smartphone. The collected information includes the registered participants, the registered devices, and all the games played. The biometric data collected from the built-in sensors (accelerometer, gyroscope, magnetometer, and device motion sensor) from the smartphone. We use the raw behavioral biometric data collected from the embedded sensors (accelerometer, gyroscope, and magnetometer) of the smartphone. We randomly select 100 smartphone users' behavioral biometric data from the BrainRun dataset to perform the experiments.

\textbf{IDNet dataset:} IDNet dataset \cite{idnet} is a gait data set from 50 participants, which is collected from inertial sensors (accelerometer, gyroscope, and magnetometer, etc.). Participants are asked to put their smartphones in the right front pocket of their trousers during sensor data collection phase. To bring experimental scenarios closer to real world scenarios, participants are asked to walk as they felt comfortable during behavioral biometrics collection. We use behavioral biometrics collected from three inertial sensors (accelerometer, gyroscope, magnetometer) to perform experiments.

\subsection{Performance on Different Public Datasets}

\begin{table}\centering
	\caption{The mean values of FAR (\%), FRR (\%), EER (\%), and AUROC on different datasets.}
	\label{performance on different datasets}
	\begin{tabular}{lcccc}
		\toprule
		Dataset&FAR $\downarrow$&FRR $\downarrow$&EER $\downarrow$&AUROC $\uparrow$\\
		\midrule
		HMOG&0.77&1.39&1.05&0.998\\
		BrainRun&0.99&2.05&1.09&0.997\\
		IDNet&0.94&1.52&1.08&0.997\\
		\bottomrule
	\end{tabular}
\end{table} 


\begin{table}[!t]
	\centering
	\caption{The performance of ten randomly selected users from different public datasets.}
	\label{detailed performance}
	\begin{tabular}{lccccc}
		\toprule
		Dataset&User ID&FAR $\downarrow$&FRR $\downarrow$&EER $\downarrow$&AUROC $\uparrow$\\
		\midrule
		\multirow{10}{*}{HMOG}
		&100669&0.32&1.52&0.69&0.998\\
		&180679&0.61&1.47&1.13&0.998\\
		&220962&0.97&1.50&1.25&0.997\\
		&352716&0.46&1.39&0.99&0.998\\
		&525584&1.24&1.47&1.37&0.997\\
		&553321&0.71&1.51&1.03&0.998\\
		&622852&0.02&1.00&0.34&0.999\\
		&745224&0.79&1.52&1.21&0.998\\
		&799296&1.46&1.49&1.46&0.997\\
		&962159&0.43&1.49&1.03&0.998\\
		\midrule
		\multirow{10}{*}{BrainRun}
		&6jtbpdh&0.24&2.00&0.25&0.999\\
		&8xjh8a&0.98&1.69&0.98&0.998\\
		&9gx7uks&1.34&1.52&1.36&0.997\\
		&gzx7rv&0.18&1.67&0.20&0.998\\
		&sxvkh3b&0.86&1.67&0.88&0.998\\
		&uui53he&1.09&2.56&1.28&0.997\\
		&ioxyr9y&0.87&2.00&0.95&0.998\\
		&w8f2wrs&0.97&1.81&1.10&0.998\\
		&68n9ll&0.77&1.83&0.92&0.998\\
		&d99p79w&2.32&2.60&2.32&0.992\\
		\midrule
		\multirow{10}{*}{IDNet}
		&u002&0.89&1.15&0.96&0.998\\
		&u004&1.65&1.60&1.61&0.996\\
		&u008&1.55&1.89&1.55&0.997\\
		&u012&1.35&1.61&1.48&0.997\\
		&u013&0.49&1.58&0.79&0.998\\
		&u024&0.18&1.19&0.31&0.999\\
		&u029&0.89&1.72&0.93&0.998\\
		&u041&0.86&1.32&0.88&0.998\\
		&u043&1.47&1.66&1.47&0.997\\
		&u046&1.29&1.55&1.29&0.997\\
		\bottomrule
	\end{tabular}
\end{table}

To evaluate the performance of the proposed relative attention-based one-class adversarial autoencoder for the continuous authentication task, we perform several experiments on three public datasets. For each experiment, we randomly select one user from \texttt{N} users as the legitimate user and the rest \texttt{N-1} as impostors. In the training phase, we randomly choose 80\% samples of the legitimate user to train the relative attention-based one-class adversarial autoencoder. In the test phase, the rest 20\% samples of the legitimate user and the whole samples of the impostors are used to test the trained authentication model. We train an authentication model for each smartphone user. Table \ref{performance on different datasets} lists the mean FAR, FRR, EER, and AUROC of all smartphone users for each dataset. As shown in Table \ref{performance on different datasets}, the proposed relative attention-based one-class adversarial autoencoder achieves excellent performance on three public datasets. The proposed authentication method achieves 0.77\% FAR, 1.39\% FRR, 1.05\% EER and 0.998 AUROC on the HMOG dataset, and 0.99\% FAR, 2.05\% FRR, 1.09\% EER and 0.997 AUROC on the BrainRun dataset. It also reaches an average of 0.94\% FAR, 1.52\% FRR, 1.08\% EER, and 0.997 AUROC on the IDNet dataset. Besides, to be able to observe the authentication performance of the proposed relative attention-based one-class adversarial autoencoder on single smartphone user, we also list the FAR, FRR, EER, and AUROC of ten randomly selected smartphone users from different public datasets in Table \ref{detailed performance}. As shown in Table \ref{detailed performance}, the proposed authentication method can achieve excellent FAR (less than 2.5\%), FRR (less than 3\%), EER (less than 2.5\%), and AUROC (more than 0.990) on each random user. 

\begin{figure}[!t]
	\centering
	
	\includegraphics[width=0.95\linewidth]{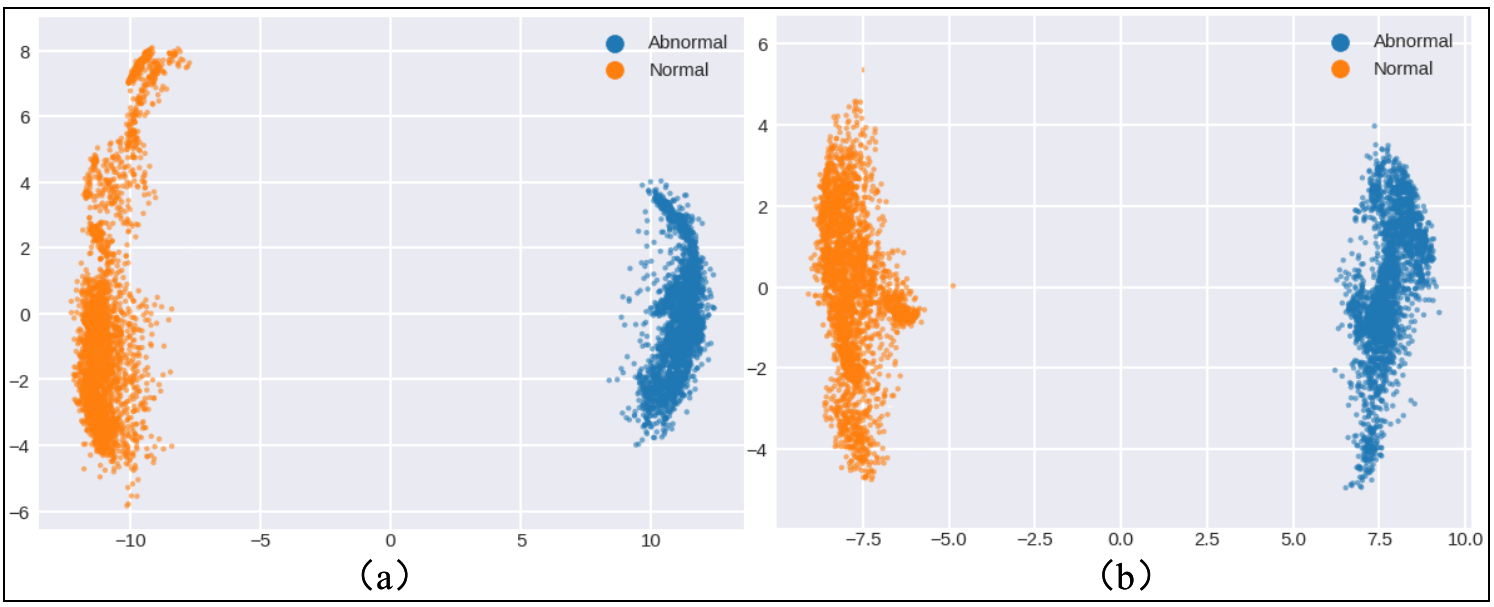}
	
	\caption{Extracted deep features visualization of two randomly selected subjects on the HMOG dataset. The deep features are extracted from the encoder of the proposed relative attention-based one-class adversarial autoencoder, and a reduced dimension of 2 by the PCA. The normal samples plotted with orange are from the test subject, and the abnormal samples plotted with blue are from the remaining smartphone users.}
	\label{Feature visualization}
\end{figure}

Besides, we conducted deep abstract semantic features visualization experiments on the HMOG dataset to demonstrate whether the proposed relative attention-based one-class adversarial autoencoder can effectively learn rich contextual semantic information that represents legitimate users' behavioral patterns. As shown in Fig. \ref{Feature visualization}, the extracted deep features from the encoder of the proposed relative attention-based one-class adversarial autoencoder are able to distinguish between the normal samples and abnormal samples.

\subsection{Comparison with Representative Continuous Authentication Methods}
Current continuous authentication methods are based on different datasets for performance evaluation. To make performance comparison experiments more convincing, we reproduce these authentication methods on three common datasets. Besides, we also make a qualitative comparison with representative continuous authentication work. The Acc., Gyr., Mag., To., Ori, El., and Gra. indicate the Accelerometer, Gyroscope, Magnetometer, Touch, Orientation, Elevation, and Gravity, respectively.

\begin{table*}\centering
	\caption{The EER (\%) of representative continuous authentication methods on three datasets.}
	\label{representative work}
		\begin{tabular}{lccccc}
			\toprule
			\multirow{2}{*}{Work}&\multirow{2}{*}{Classifiers}&\multirow{2}{*}{Training data}& \multicolumn{3}{c}{Dataset} \\
			\cmidrule{4-6}
			&&&HMOG&BrainRun&IDNet\\
			\midrule
			Roy et al. (2015) \cite{roy2015hmm}&HMM&Owner&10.21&13.15&14.29\\
			Sitová et al. (2016) \cite{sitova2016hmog}&Scaled Manhattan&Owner&16.34&18.64&19.07\\
			Centeno et al. (2018) \cite{centeno2018mobile}&OC-SVM&Owner \& Impostors&3.51&4.27&4.82\\
			Shen et al. (2017) \cite{shen2017performance}&HMM&Owner&8.46&10.65&10.79\\
			Li et al. (2020) \cite{li2020scanet}&OC-SVM&Owner \& Impostor&5.85&7.21&7.93\\
			Cherifi et al. (2021) \cite{cherifi2021efficient} & HMM&Owner &15.49&17.86&18.31\\
			Wang et al. (2021) \cite{wang2021framework} & Deep Metric Learning&Owner \& Impostor&4.13&5.21&6.97\\
			Shen et al. (2022) \cite{shen2022mmauth}&DeSVDD&Owner&6.97&5.63&11.15\\
			Hu et al. (2023) \cite{hu2023authconformer}&OC-SVM&Owner \& Impostor&1.06&1.25&1.19\\
			Li et al. (2024) \cite{li2024snnauth}&Spiking Neural Networks&Owner&2.19&3.72&4.13\\
			Yang et al. (2024) \cite{yang2024unsupervised}&One-Dimensional Autoencoder&Owner&4.27&5.49&6.31\\
			Hu et al. (2025) \cite{hu2025behavioral}&Masked Autoencoder&Owner&0.68&0.94&2.14\\
			Our work & Adversarial Autoencoder&Owner&1.05&1.09&1.08\\
			\bottomrule
		\end{tabular}
\end{table*} 

\begin{table*}\centering
	\caption{Qualitative comparison with representative continuous authentication methods.}
	\label{qualitative comparison}
	\resizebox{\linewidth}{!}{
		\begin{tabular}{lccccc}
			\toprule
			Work&Sensors &Classifier&Training data& Results&Time (s) \\
			\midrule
			Sitová et al. (2016) \cite{sitova2016hmog}&Acc., Gyr., Mag., To.&Scaled Manhattan&Owner&EER: 7.16\% Walking&~60\\
			Centeno et al. (2018) \cite{centeno2018mobile}&Acc., Gyr., Mag.&OC-SVM&Owner \& Impostor&Accuracy: 97.8\%&\textgreater1\\
			Shen et al. (2017) \cite{shen2017performance}&Acc., Gyr., Mag., Ori.&HMM&Owner&EER: 4.74\%&8\\
			Yang et al. (2019) \cite{yang2019behavesense}&To.&OC-SVM&Owner&Accuracy: 95.85\%&0.01\\
			Abuhamad et al. (2020) \cite{abuhamad2020autosen}&Acc., Gyr., Mag., To., El.&LSTM&Owner \& Impostor&FAR: 0.96\%, FRR: 8.08\%& \textgreater0.5\\
			Li et al. (2020) \cite{li2020scanet}&Acc., Gyr.&OC-SVM&Owner \& Impostor&EER: 5.14\%&3\\
			Zhu et al. (2020) \cite{zhu2020espialcog}&Acc., Gyr., Gra.&LSTM&Owner \& Impostor&Accuracy: 91.59\%& \textgreater3\\
			Li et al. (2021) \cite{li2021cnn}&Acc., Gyr., Mag.&Isolation Forest&Owner \& Impostor&EER: 3.64\%&\textgreater2\\
			Wang et al. (2021) \cite{wang2021framework}&Acc.&CNN&Owner \& Impostor&Accuracy: 95.3\% Mcgill&—\\
			Ray-Dowling et al. (2022) \cite{ray2022evaluating}&Multi-Modality&Binary SVM&Owner \& Impostor&EER: 1.5\% HMOG, 0.2\% BB-MAS&\textless38.5\\
			Shen et al. (2022) \cite{shen2022mmauth}&Acc., Gyr., Mag., To., etc.&DeSVDD&Owner&EER: 14.9\%&—\\
			Li et al. (2024) \cite{li2024snnauth}&Acc., Gyr., Mag.&Spiking Neural Networks&Owner \& Impostor&EER: 1.85\%&—\\
			Yang et al. (2024) \cite{yang2024unsupervised}&Acc., Gyr., Mag.&One-Dimensional Autoencoder&Owner&EER: 3.86\%&—\\ 
			Our work &Acc., Gyr., Mag.& Autoencoder&Owner &EER: 1.05\% HMOG&0.7\\
			\bottomrule
		\end{tabular}
	}
\end{table*} 


\textbf{Performance comparison on three common datasets:} We reproduce twelve representative continuous authentication methods on the HMOG dataset, BrainRun dataset, and IDNet dataset, including Roy et al. \cite{roy2015hmm}, Sitová et al. \cite{sitova2016hmog}, Centeno et al. \cite{centeno2018mobile}, Shen et al. \cite{shen2017performance}, Li et al. \cite{li2020scanet}, Cherifi et al. \cite{cherifi2021efficient}, Wang et al. \cite{wang2021framework}, Shen et al. \cite{shen2022mmauth}, Hu et al. \cite{hu2023authconformer}, Li et al. \cite{li2024snnauth}, Yang et al. \cite{yang2024unsupervised}, and Hu et al. \cite{hu2025behavioral}. During the experimental reproduce phase, we obtain the experimental settings from these articles. The traditional machine learning classification algorithms are from the scikit-learn library: one-class support vector machine (OC-SVM), hidden markov model (HMM). The deep metric learning models in \cite{centeno2018mobile} and \cite{wang2021framework} are implemented with the deep learning framework Keras, and the deep learning models in \cite{li2020scanet}, \cite{shen2022mmauth}, \cite{hu2023authconformer}, \cite{li2024snnauth}, \cite{yang2024unsupervised}, and \cite{hu2025behavioral} are implemented with the PyTorch framework. Table \ref{representative work} lists the Classifier, the training data, and the EER of representative work on three common datasets. As shown in Table \ref{representative work}, the baseline method \cite{hu2025behavioral} achieves competitive authentication performance among twelve baselines. However, it still encounters two limitations: 1) It needs to train two separate models, the masked latent representation generator and the reconstructor. Maintaining and optimizing two separate models increases the complexity of the training process. This complexity can also complicate hyperparameter tuning, as the interaction between the two models may necessitate careful coordination to achieve optimal performance; 2) It needs to empirically select an appropriate masking ratio or pattern to improve the model's ability to learn meaningful representations. Among these representative authentication methods, our authentication approach achieves the best EER with 1.05\%, 1.09\%, and 1.08\% on three datasets, respectively. More importantly, compared with the baseline method \cite{hu2025behavioral}, the proposed relative attention-based one-class autoencoder, along with the latent and sample discriminators, are trained and optimized in a unified framework, eliminating the need for manual parameter adjustment and intervention during the training process.

\textbf{Qualitative comparison with representative continuous authentication work:} We also make a qualitative comparison with representative continuous authentication work, including Sitová et al. \cite{sitova2016hmog}, Centeno et al. \cite{centeno2018mobile}, Shen et al. \cite{shen2017performance}, Yang et al. \cite{yang2019behavesense}, Abuhamad et al. \cite{abuhamad2020autosen}, Li et al. \cite{li2020scanet}, Zhu et al. \cite{zhu2020espialcog}, Li et al. \cite{li2021cnn}, Wang et al. \cite{wang2021framework}, Ray-Dowling et al. \cite{ray2022evaluating}, Shen et al. \cite{shen2022mmauth}, Li et al. \cite{li2024snnauth}, and Yang et al. \cite{yang2024unsupervised}. 

Table \ref{qualitative comparison} lists the sensors, classifiers, training data, experimental result, and time overhead of each representative work. Sitová et al. \cite{sitova2016hmog} introduced a public dataset of smartphone users' behavioral biometric features for the continuous authentication task. Experimental results shown that they can achieve the best authentication performance (7.16\% EER in walking and 10.05\% in sitting) when they fuse HMOG, taps, and keystroke features. Centeno et al. \cite{centeno2018mobile} apply a siamese convolutional neural network to extract deep features from sensor data signals, and they achieve 97.8\% accuracy on the HMOG dataset using the one-class SVM classifier. Shen et al. \cite{shen2017performance} explored the contribution of each motion sensor behavior for the continuous authentication performance. Experimental results shown that they achieve the lowest EER of 4.74\% when they combine the accelerometer, gyroscope, magnetometer, and orientation sensors. Yang et al. \cite{yang2019behavesense} explored the continuous authentication performance of different type of screen touch operations, including click operations and sliding operations. Abuhamad et al. \cite{abuhamad2020autosen} proposed an LSTM-based end-to-end continuous authentication approach, and explored different LSTM architectures in learning and capturing the behavioral patterns of smartphone users. Experimental results shown that they can achieve 0.96\% FAR and 8.08\% FRR using readings of only three sensors. Li et al. \cite{li2020scanet} utilize a two-stream convolutional neural network architecture to extract deep features from behavioral biometrics, and the OC-SVM classifier is used to classify the access user as a legitimate user or an impostor. Experimental results shown that the proposed authentication method can achieve an mean EER of 5.14\% with approximately 3s authentication time. Zhu et al. \cite{zhu2020espialcog} applied an optimized LSTM architecture to learn the behavioral patterns from three built-in sensors. Besides, they evaluated the performance of the proposed continuous authentication method on a large-scale real-world noisy dataset. Li et al. \cite{li2021cnn} apply a conditional wasserstein generative adversarial network to generate additional sensor data signals for data augmentation, and a convolutional neural network architecture is used to extract deep features from sensor data signals. Experimental results shown that they achieve the lowest EER of 3.64\% using three motion sensors. Wang et al. \cite{wang2021framework} proposed a deep metric learning-based continuous authentication framework for smartphone users, which can be trained on the battery-powered smartphone. Experimental results demonstrate that they achieve authentication accuracy over 95\% using only one motion sensor. Ray-Dowling et al. \cite{ray2022evaluating} evaluated the authentication performance of individual behavioral biometrics modalities as well as their fusion on two public mobile datasets, the HMOG dataset and BB-MAS. The experimental results demonstrated that the best performance was achieved with an EER of 1.5\% for the HMOG dataset and 0.2\% for the BB-MAS dataset when using the binary SVM. Shen et al. \cite{shen2022mmauth} constructed a smartphone users' behavioral biometrics dataset for the continuous authentication task, which are collected based on unconstrained touch operations from 100 volunteers. Experimental results demonstrate that they can achieve 14.9\% EER using a deep learning based one-class classifier (DeSVDD) on their dataset. Li et al. \cite{li2024snnauth} utilize Spiking Neural Networks (SNNs) to capture smartphone users' behavioral patterns captured from behavioral biometric data. Yang et al. \cite{yang2024unsupervised} proposed an unsupervised multisensor-based continuous authentication system with a low-rank transformer using learning-to-rank algorithms. Different from these representative works, we propose a relative attention-based one-class adversarial autoencoder to authenticate the smartphone user, which can achieve 1.05\% EER with a high authentication frequency (0.7s) on the HMOG dataset.

\textit{Remark}: The lack of a generic dataset as a benchmark to evaluate the performance of continuous authentication methods has been a limitation for behavioral biometrics-based continuous authentication of smartphone users. Most researchers evaluated their behavioral biometrics-based continuous authentication approaches on their own datasets. For a fair performance evaluation comparison, as shown in Table \ref{representative work}, we reproduce twelve representative continuous authentication approaches on three public datasets. Besides, due to the lack of technical details and detailed implementation steps of some representative authentication methods, we cannot reproduce them, we can only make a qualitative comparison with some representative continuous authentication methods in Table \ref{qualitative comparison}. Lack of a fair performance comparison on the generic dataset is also a common problem with current behavioral biometrics-based authentication methods.

\begin{table*}\centering
	\caption{Ablation study of the proposed relative attention-based one-class adversarial autoencoder on three datasets.}
	\label{Ablation study performance on different datasets}
	\begin{tabular}{lccccc}
		\toprule
		Architecture&Dataset&FAR $\downarrow$&FRR $\downarrow$&EER $\downarrow$\\
		\midrule
		\multirow{3}{*}{Without two discriminators and relative attention layer}&HMOG&4.37&2.81&3.58\\
		&BrainRun&4.64&2.97&3.83\\
		&IDNet&4.76&2.85&3.79\\
		\midrule
		\multirow{3}{*}{With relative attention layer}&HMOG&1.69&2.22&1.83\\
		&BrainRun&1.85&2.45&1.91\\
		&IDNet&1.92&2.31&1.88\\
		\midrule
		\multirow{3}{*}{With two discriminators}&HMOG&2.98&2.55&2.74\\
		&BrainRun&3.23&2.71&2.98\\
		&IDNet&3.37&2.65&2.84\\
		\midrule
		\multirow{3}{*}{Two discriminators + relative attention layer}&HMOG&0.77&1.39&1.05\\
		&BrainRun&0.99&2.05&1.09\\
		&IDNet&0.94&1.52&1.08\\
		\bottomrule
	\end{tabular}
\end{table*} 

\subsection{Ablation Study}

In order to evaluate the effectiveness of each part (relative attention layer, and two discriminators) in proposed one-class adversarial autoencoder architecture, we carry ablation study on three datasets (HMOG dataset, BrainRun dataset, and IDNet dataset). We remove the relative attention layer, and two discriminators from the proposed one-class adversarial autoencoder architecture, and perform authentication performance evaluation experiments on three datasets, respectively. Table \ref{Ablation study performance on different datasets} lists the mean experimental results without each part on three datasets. As shown in Table \ref{Ablation study performance on different datasets}, the relative attention layer play an important role in improving authentication performance, the authentication performance of EER improves further by 1.5\% with the relative attention layers. Experimental results demonstrate that we can further improve the authentication performance through the effective stacking of convolutional layers and relative attention layers. Besides, when the two discriminators are added, the authentication performance of EER is improved further by a 0.8\% on three public datasets.

\begin{figure*}[!t]
	\centering
	
	\includegraphics[width=0.60\linewidth]{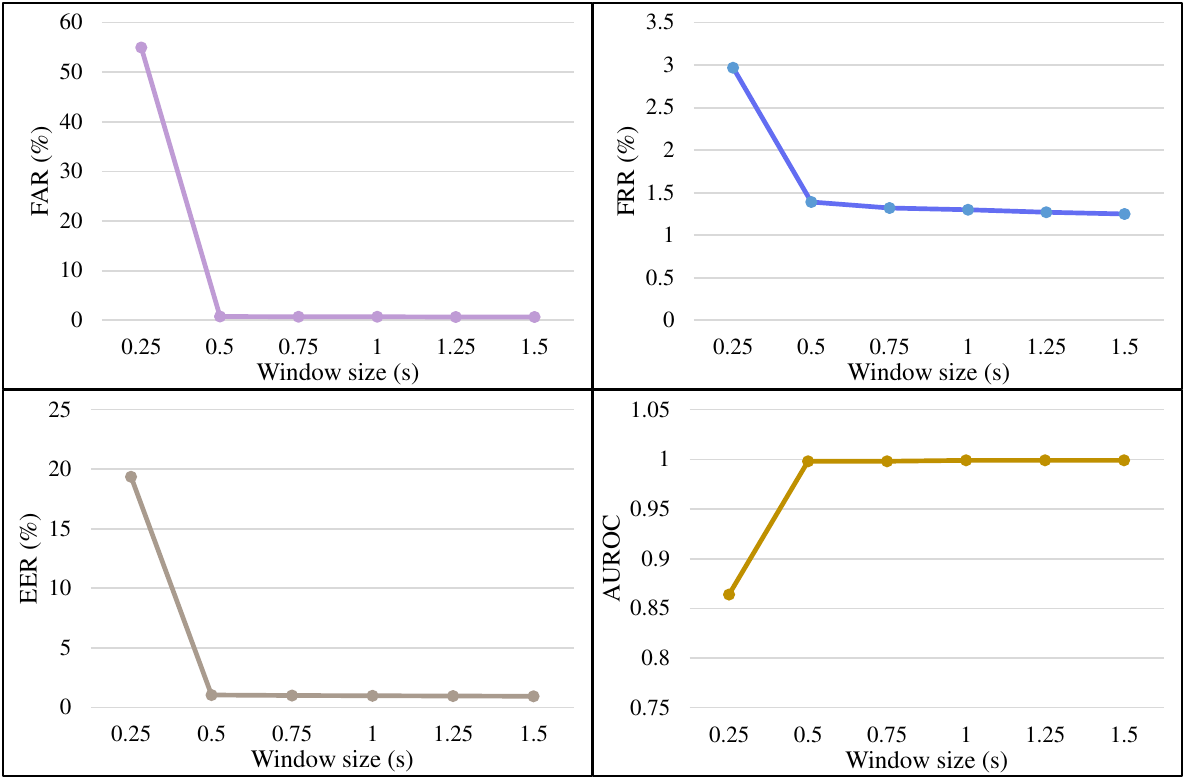}
	
	\caption{The trend of the FAR, FRR, EER, and AUROC with the increase of time window size.}
	\label{window size trend}
\end{figure*} 

\begin{table}\centering
	\caption{The mean values of FAR (\%), FRR (\%), EER (\%), and AUROC with different time window size.}
	\label{window size}
	\begin{tabular}{lcccc}
		\toprule
		Window size (s)&FAR $\downarrow$&FRR $\downarrow$&EER $\downarrow$&AUROC $\uparrow$\\
		\midrule
		0.25&54.97&2.97&19.36&0.864\\
		0.50&0.77&1.39&1.05&0.998\\
		0.75&0.72&1.32&1.01&0.998\\
		1&0.69&1.30&0.98&0.999\\
		1.25&0.68&1.27&0.96&0.999\\
		1.50&0.65&1.25&0.94&0.999\\
		\bottomrule
	\end{tabular}
\end{table} 

\subsection{Performance with Different Time Windows}

To evaluate the authentication performance with different time window size of behavioral biometrics sequences, we perform several experiments with different time window size on the HMOG dataset. Fig. \ref{window size trend} shows the trend of the FAR, FRR, EER, and AUROC with the increase of time window size. The FAR in Fig. \ref{window size trend}a decreases with the increase of time window size from 0.25s to 0.5s, and then fluctuates slightly with the increase of time window size from 0.5s to 1.5s. The FRR and EER in Fig. \ref{window size trend}b and Fig. \ref{window size trend}c show the trend to the FAR. The AUROC in Fig. \ref{window size trend}d increases with the increase of time window size from 0.25s to 0.5s, and then keeps stable with the increase of time window size from 0.5s to 1.5s. Table \ref{window size} lists the mean values of the FAR, FRR, EER, and AUROC with different time window size. As shown in Table \ref{window size}, we achieve excellent performance with the increase of time window size from 0.5s to 1.5s. In practice, we need to balance factors such as authentication performance, authentication speed, and authentication overhead when selecting the time window size of behavioral biometric sequences. If the window size is too small, the proposed continuous authentication model cannot effectively learn users' behavioral patterns, resulting in unsatisfactory authentication performance. If the window size is too large, attackers cannot be identified in a timely manner, and the authentication overhead is relatively high. Considering the time cost and authentication efficiency, we choose a 0.5s time window size of behavioral biometrics sequences in the experiments.

\subsection{Anti-Attack Analysis}

To evaluate the robustness of the proposed continuous authentication system in a real-world context, this study examines two types of attacks: random attacks and impersonation attacks.

\textbf{Random attack}: Random attack means that the attacker attempts to attack the legitimate user's smartphone with his behavioral habit, he has no knowledge of the legitimate user’s behavioral patterns before attacking. In the experiments, similar to Wang et al. \cite{wang2021framework}, we choose the smartphone user from HMOG dataset as the legitimate user, and the smartphone users from the BrainRun and IDNet dataset as attackers. Since different datasets are the behavioral biometrics of smartphone users collected under different scenarios, they can be used to simulate random attacks by multi-attackers. For each experiment, we randomly choose one smartphone user's behavioral biometrics from HMOG dataset (100 users) to train the relative attention-based one-class adversarial autoencoder, and the behavioral biometrics from the BrainRun and IDNet dataset are used to launch attacks. We perform random attacks analysis on each smartphone user in the HMOG dataset (100 users).

\textbf{Impersonation Attack:} Before launching an attack, the attacker attempts to imitate behavioral patterns of the legitimate user by observing his behavioral habits during interaction with the smartphone. In this paper, we invited 20 volunteers to participate in the impersonation attack experiments. They signed an agreement to collect their behavioral biometric data for use in the impersonation attacks. We conducted 20 rounds of experiments, with each volunteer posing as a legitimate user. In each round, we randomly selected one of the 20 volunteers to act as the legitimate user, while the remaining 19 participants acted as attackers to conduct the impersonation attacks. In the impersonation attack experiments, the attacker first observed the legitimate users' behavioral patterns while interacting with their smartphones. Then attackers try to interact with legitimate users’ smartphones through the learned behavioral patterns.

Table \ref{random attack} lists the mean values of FAR, FRR, and EER, under random attack and impersonation attack. As shown in Table \ref{random attack}, the proposed authentication method can achieve 0.01\% FAR, 1.41\% FRR, 1.28\% EER, 0.996 AUROC, and 0.01\% FAR, 1.35\% FRR, 1.23\% EER under the random attacks from the BrainRun dataset and IDNet dataset, respectively. More importantly, we can achieve the mean performance of 1.63\% FAR, 1.45\% FRR, and 1.51\% EER under the impersonation attack. Experimental results demonstrate that the proposed continuous authentication approach can well resist random attacks and impersonation attacks.

\begin{table}\centering
	\caption{The mean values of FAR (\%), FRR (\%), and EER (\%) under random attack and impersonation attack.}
	\label{random attack}
	\begin{tabular}{lcccc}
		\toprule
		Attacker&FAR $\downarrow$&FRR $\downarrow$&EER $\downarrow$\\
		\midrule
		BrainRun&0.01&1.41&1.28\\
		IDNet&0.01&1.35&1.23\\
		Imitators&1.63&1.45&1.51\\
		\bottomrule
	\end{tabular}
\end{table} 

\subsection{Overhead Analysis}

We evaluate the resource consumption of the proposed relative attention-based one-class adversarial autoencoder for continuous authentication of smartphone users in three aspects, including time efficiency, storage, and model parameter size and FLOPs.

\textbf{Time efficiency:} The time cost of the proposed continuous authentication method consists of three parts, including a certain time window of sensor data sequence for the authentication (t1), the time of data preprocessing (t2), and the time of the proposed relative attention-based one-class adversarial autoencoder for authentication (t3). In the experiments, we choose a 0.5s time window of sensor data sequence for the authentication (t1=0.5s). The time of data preprocessing is 130ms (t2=130ms), including noise removing, data normalization, and sample generation. The time of the proposed relative attention-based one-class adversarial autoencoder for the authentication is 100ms (t3=100ms). Then the total time cost is approximately 0.7s (t1+t2+t3).

\textbf{Storage:} The pretrained relative attention-based one-class adversarial autoencoder consists of four parts, including the encoder, the decoder, the latent space discriminator, and the sample discriminator. The encoder and decoder have size of 650.9KB and 574.6KB, respectively. The latent space discriminator and sample discriminator have size of 970.1KB and 630.8KB, respectively. Then the total size of the proposed relative attention-based one-class adversarial autoencoder is 2.7MB.

\textbf{Model parameter size and FLOPs:} The encoder has 0.15M parameters and 0.01G FLOPs. The decoder has 0.1M parameters and 0.01G FLOPs. The latent space discriminator has 0.23M parameters. The sample discriminator has 0.15M parameters and 0.09G FLOPs. Then the proposed relative attention-based one-class adversarial autoencoder has 0.63M parameters and 0.11G FLOPs in total.

\section{Conclusion}
\label{conclusion}

In this work, we propose a relative attention-based one-class adversarial autoencoder for continuous authentication of smartphone users. We combine the convolutional layers and the constructed relative attention layers to capture rich contextual semantic representation of smartphone user's behavioral patterns. More importantly, the proposed relative attention-based one-class adversarial autoencoder only learns the latent representations of legitimate user's behavioral patterns, which is trained without negative samples from impostors. The proposed authentication method achieves excellent performance with a high authentication frequency (0.7s) on three public dataset. Experimental results show that the proposed authentication method achieves 0.77\% FAR, 1.39\% FRR, 1.05\% EER and 0.998 AUROC on the HMOG dataset. We also achieve excellent performance with 0.99\% FAR, 2.05\% FRR, 1.09\% EER and 0.997 AUROC on the BrainRun dataset, and 0.94\% FAR, 1.52\% FRR, 1.08\% EER and 0.997 AUROC on the IDNet dataset. Compared with representative continuous authenticaton methods, experimental results demonstrate that the proposed authentication method can achieve superior authentication performance. Besides, experimental results show that the proposed authentication method can achieve 0.01\% FAR, 1.41\% FRR, 1.28\% EER, 0.996 AUROC, and 0.01\% FAR, 1.35\% FRR, 1.23\% EER, 0.997 AUROC, when the behavioral biometrics from BrainRun dataset and IDNet dataset are used to launch the random attacks, respectively. Experimental results also demonstrate that we can achieve 1.63\% FAR, 1.45\% FRR, and 1.51\% EER under the impersonation attack.

\bibliographystyle{IEEEtran}
\bibliography{egbib}

\end{document}